\documentclass[prl,aps,twocolumn,amsfonts,showpacs,superscriptaddress]{revtex4}

\usepackage{epsfig} 
\usepackage{psfrag}
\usepackage{amsmath}
\usepackage{amssymb}
\usepackage{color}
\usepackage{graphicx}

\begin{document}

\title{Polaritons and Pairing Phenomena in Bose--Hubbard Mixtures}
\author{M. J. Bhaseen} 
\affiliation{University of Cambridge, Cavendish
Laboratory, Cambridge, CB3 0HE, UK.}  
\author{M. Hohenadler}
\affiliation{University of Cambridge, Cavendish Laboratory, Cambridge,
CB3 0HE, UK.}  
\affiliation{Present Address: OSRAM Opto Semiconductors
GmbH, 93055 Regensburg, GER.}  
\author{A. O. Silver}
\affiliation{University of Cambridge, Cavendish Laboratory, Cambridge,
CB3 0HE, UK.}  \author{B. D. Simons} 
\affiliation{University of
Cambridge, Cavendish Laboratory, Cambridge, CB3 0HE, UK.}
\date{\today}

\begin{abstract}
Motivated by recent experiments on cold atomic gases in ultra high
finesse optical cavities, we consider the two-band Bose--Hubbard model
coupled to quantum light. Photoexcitation promotes carriers between
the bands and we study the interplay between Mott insulating behavior
and superfluidity. The model displays a ${\rm U}(1)\times {\rm U}(1)$
symmetry which supports the coexistence of Mott insulating and
superfluid phases, and yields a rich phase diagram with multicritical
points. This symmetry is shared by several other problems of current
experimental interest, including two-component Bose gases in optical
lattices, and the bosonic BEC-BCS crossover for atom-molecule mixtures
induced by a Feshbach resonance. We corroborate our findings by
numerical simulations.
\end{abstract}


\maketitle

{\em Introduction.}--- The spectacular advances in cold atomic gases
have led to landmark experiments in strongly correlated systems.  With
the observation of the superfluid--Mott insulator transition in
$^{87}{\rm Rb}$~\cite{Greiner:SI}, and the BEC--BCS crossover in
$^{40}{\rm K}$~\cite{Regal:BCSBEC}, attention is now being directed
towards multicomponent gases. Whether they be distinct atoms or
internal states, such systems bring ``isospin'' degrees of freedom.
They offer the fascinating prospect to realize novel phases, and to
study quantum magnetism, Mott physics and
superfluidity~\cite{Altman:Twocomp}.

More recently, significant experimental progress has been made in
combining cavity quantum electrodynamics (cavity QED) with ultracold
gases. Strong atom-field coupling has been achieved using ultra high finesse
optical cavities \cite{Brennecke:Cavity}, and with optical fibres 
on atom chips \cite{Colombe:Strong}. These experiments
open an exciting new chapter in coherent matter--light interaction, and 
have already led to pioneering studies of condensate dynamics \cite{Brennecke:Opto}.  
The light field serves not only as a probe of the many--body system, but 
may also support
interesting cavity mediated phenomena and phases. This dual role 
has been exploited in studies of polariton 
condensates in semiconductor microcavities \cite{Kasprzak:BEC}. 
It is reinforced by ground breaking cavity QED experiments
using superconducting qubits in microwave resonators \cite{Wallraff:Strong}. This has led to solid state 
measurements of the collective states of the Dicke model 
\cite{Wallraff:Collective}, and remarkable observations of the 
Lamb shift \cite{Wallraff:Lamb}. 

In this work we examine the impact of cavity radiation on the
Bose--Hubbard model. We focus on a two-band model in which photons
induce transitions between two internal states or Bloch bands. This is
a natural generalization of the much studied two-level systems coupled
to radiation, and may serve as a useful paradigm in other
contexts. The new ingredients are that the bosonic carriers may form a
Mott insulator, or indeed condense. The primary question is whether a
novel Mott insulating state can survive, which supports a condensate
of photoexcitations or mobile defects.  In analogy with zero point
motion in Helium~\cite{Andreev:SS}, this may be viewed as a form of
supersolid in which fluctuations of the photon field induce
defects. Whilst this question has its origin in polariton condensates
in fermionic band insulators~\cite{Littlewood:Models}, the present
problem is rather different. Since the integrity of the Mott state is
tied to the interactions, {\em a priori} it is unclear that it
survives the effects of itinerancy and photoexcitation.  Nonetheless,
the outcome is affirmative, and the model displays both this novel
phase and a rich phase diagram. Related phases were recently 
observed in simulations of other two-component
models~\cite{Soyler:Twocomp,Rousseau:Fesh}.

{\em The Model.}--- Let us consider a two-band Bose--Hubbard model
coupled to the light field of an optical cavity within the rotating
wave approximation
\begin{equation}
\begin{aligned}
H_0 & =\sum_{i\alpha} \epsilon_\alpha n_i^\alpha
+\sum_{i\alpha}\frac{U_{\alpha}}{2}n_i^\alpha(n_i^\alpha-1)
+V\sum_i n_i^an_i^b\\
& -\sum_{\langle ij\rangle\alpha} J_\alpha \alpha_i^\dagger \alpha_j
+\omega \psi^\dagger \psi+g\sum_i\left(b_i^\dagger a_i\psi
+{\rm h.c.}\right),
\end{aligned}
\label{Hdef}
\vspace{-0.3cm}
\end{equation}
where $\alpha=a,b$ are two bands of bosons with $[\alpha_i,\alpha_j^\dagger]=\delta_{ij}$. These might be states of
different orbital or spin angular momentum. Here, $\epsilon_\alpha$,
effects the band splitting, $U_\alpha$ and $V$ are
interactions, $J_\alpha$, are nearest--neighbor hopping parameters,
and $\omega$ is the frequency of the mode, $\psi$. We
consider just a single mode, which couples uniformly to the
bands. The coupling, $g$, is the strength of the matter--light
interaction. In view of the box normalization of the photon, we
denote $g\equiv\bar g/\sqrt{N}$, where $N$ is the number of lattice
sites.  It is readily seen that $N_1=\sum_i (n_i^b+n_i^a)$ and
$N_2=\psi^\dagger\psi+\sum_i(n_i^b-n_i^a+1)/2$, commute with $H_0$. 
These are the total number of atomic carriers, and
photoexcitations (or polaritons) respectively. These conservation laws
reflect the global ${\rm U}(1)\times {\rm U}(1)$ symmetry of 
$H_0$, such that $a\rightarrow e^{i\vartheta}a$,
$b\rightarrow e^{i\varphi} b$, $\psi\rightarrow
e^{-i(\vartheta-\varphi)}\psi$, where $\vartheta,\varphi$ are
arbitrary. This symmetry will have a direct manifestation in the phase
diagram, and suggests implications for other multicomponent
problems. We begin by assuming that $a$ are strongly
interacting hardcore bosons, and that $b$ are
dilute so that we may neglect their interactions. 

\begin{figure}
\psfrag{m1}{$\mu_1$}
\psfrag{m2}{$\mu_2$}
\psfrag{MI}{${\rm MI}$}
\psfrag{SR-MI}{${\rm SR MI}$}
\psfrag{V}{${\rm V}$}
\psfrag{U}{${\rm U}$}
\includegraphics[width=7.5cm,clip=true]{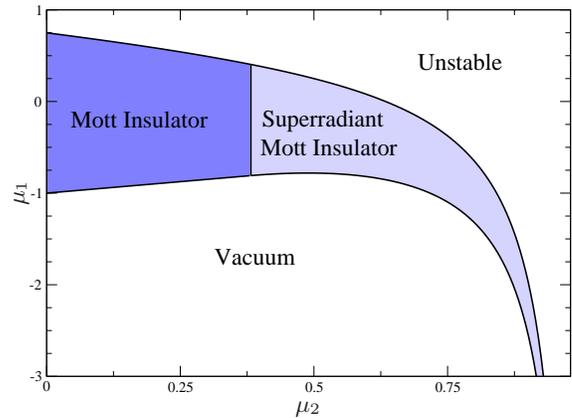}
\caption{Zero hopping phase diagram in the large-$U_a$ limit, with
$\epsilon_b=-\epsilon_a=\omega=\bar g=1$, corresponding to
$\omega<\omega_0$. The vertical line, $\bar g=\bar g_c$, is the
superradiance transition in the Dicke model, and separates a Mott
insulator with $n_a+n_b=1$ and $\langle \psi^\dagger\psi\rangle=0$
(dark blue), from a superradiant Mott insulator with
$\langle\psi^\dagger\psi\rangle\neq 0$ (light blue).  Outside are the
vacuum, and an unstable region corresponding to macroscopic population
of the $b$ states. Whilst the total density is fixed within both Mott
phases, the individual $a$ and $b$ populations vary in the
superradiant phase and may be viewed as isospin order. For
$\omega>\omega_0$, the boundaries may cross and terminate the lobe.}
\label{Fig:zerohop}
\end{figure}

{\em Zero Hopping Limit.}--- To gain insight into (\ref{Hdef}) we
examine the zero hopping limit.  This will anchor the phase diagram to
an exactly solvable many body limit. The photon couples all the sites, and in the
thermodynamic limit is described by a coherent state,
$|\gamma\rangle\equiv e^{-\frac{\gamma^2}{2}}e^{\gamma
\psi^\dagger}|0\rangle$, with mean occupation $\langle
\psi^\dagger\psi\rangle=\gamma^2$. We may replace the grand canonical
Hamiltonian, $H\equiv H_0-\mu_1N_1-\mu_2N_2$, by an effective single
site problem, $\langle \gamma |H|\gamma\rangle\equiv \sum_i {\mathcal
H}_i$:
\begin{equation}
{\mathcal H} \equiv \sum_\alpha \tilde\epsilon_\alpha n_\alpha 
+\tilde\omega\bar\gamma^2
+\bar g\bar\gamma(b^\dagger a+a^\dagger b),
\label{Heff}
\end{equation}
and we drop the offset, $-\mu_2/2$. We define $\tilde\epsilon_a\equiv
\epsilon_a-\mu_1+\mu_2/2$, $\tilde\epsilon_b\equiv
\epsilon_b-\mu_1-\mu_2/2$, $\tilde\omega\equiv\omega-\mu_2$, and the
mean photon occupation per site, $\bar\gamma^2\equiv \gamma^2/N$.  The
effective Hamiltonian (\ref{Heff}) describes a single two-level system
coupled to an effective ``radiation field'' of $b$-particles, or the
Jaynes--Cummings model; for $N$ two-level systems
this is known as the Dicke or Tavis--Cummings
model, and is integrable \cite{Dicke:Coherence,Hepp:Super}. These paradigmatic models
are well known in both atomic physics and quantum optics, and describe
localized excitons coupled to light~\cite{Littlewood:Models}. 
The eigenstates of
(\ref{Heff}) are superpositions in the upper and lower bands (that we
denote as $|0,n\rangle$ and $|1,n-1\rangle$) with total occupancy $n$.
The lowest energy is
$E_n^-=\tilde\omega\bar\gamma^2+n\tilde\epsilon_b-\tilde\omega_0/2
-\sqrt{\tilde\omega_0^2/4+\bar g^2\bar \gamma^2 n}$,
where $\tilde\omega_0\equiv \tilde\epsilon_b-\tilde\epsilon_a$.
Minimizing on $\bar\gamma$ gives a
self-consistency equation for the photon, and the resulting
eigenstates yield the zero hopping phase diagram in
Fig.~\ref{Fig:zerohop}.  In the thermodynamic limit described here, 
only the lowest Mott state, with $n_a+n_b=1$, survives;
for $\mu_1\ge \epsilon_b-\mu_2/2-\bar g^2/4\tilde\omega$ it is 
favorable to macroscopically populate the upper band.
Within this stable Mott phase the variation yields either
$\bar\gamma=0$, corresponding to zero photon occupancy, or
$\bar\gamma^2=(\bar g^4-\bar g_c^4)/4\tilde\omega^2\bar g^2$, where
$\bar g_c\equiv \sqrt{\tilde\omega\tilde\omega_0}$; the former occurs
for $\bar g\le \bar g_c$ and the latter for $\bar g\ge \bar g_c$. In
fact, this onset corresponds to the superradiance transition in the
Dicke model~\cite{Dicke:Coherence,Hepp:Super}. Indeed, since $n_a+n_b=1\equiv 2S$ in
the lowest lobe, one may construct the Dicke model directly from
(\ref{Hdef}) by using a spin $S=1/2$ Schwinger boson representation, 
where $S^+\equiv b^\dagger a$, $S^-\equiv a^\dagger b$,
$S^z\equiv (n_b-n_a)/2$:
\begin{equation}
H=\tilde\omega_0\sum_i S_i^z+\tilde\omega\psi^\dagger\psi+
g\sum_i\left(S_i^\dagger\psi+{\rm h.c.}\right).
\end{equation}
This describes $N$ two-level systems (or spins) coupled to photons,
and may be treated using {\em collective} spins, ${\bf J}\equiv
\sum_i^N {\bf S}_i$. This yields a large effective spin, which may be treated 
semiclassically as $N\rightarrow \infty$.  The onset of the
photon is accompanied by a magnetization, ${\mathcal M} \equiv \langle
J^z\rangle/N$, which also serves as an order parameter for this
continuous transition: ${\mathcal M}=-1/2$, for $\bar g\le \bar g_c$,
and ${\mathcal M}=-(\bar g_c/\bar g)^2/2$, for $\bar g\ge \bar g_c$.
This growth reflects the population imbalance, $\langle
n_b\rangle-\langle n_a\rangle$, due to photoexcitations.  The
agreement between the variational and Dicke model results is a useful
platform for departures.

\begin{figure}
\includegraphics[clip=true,width=7.5cm]{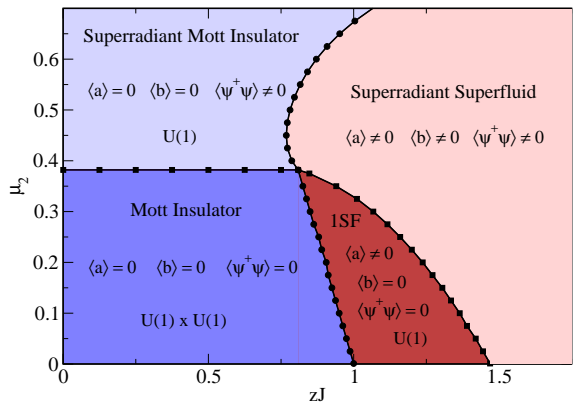}
\caption{Variational phase diagram with $J_a=J_b=J$ and
$\epsilon_a=-1$, $\epsilon_b=1$, $\omega=V=\bar g=1$, $\mu_1=0$. The
phases are (i) a Mott insulator (dark blue), (ii) a superradiant Mott state 
supporting a condensate
of photoexcitations (light blue), (iii) a superradiant superfluid 
(light red), and (iv) an
a-type superfluid (dark red).  The circles denote the transition to superfluidity as
determined by $\theta$, and the squares denote the onset of photons as
determined by $\chi$. For these parameters, the transition from the
Mott insulator to the superradiant Mott state occurs for
$\mu_2^c=(3-\sqrt{5})/2\approx 0.382$.}
\label{Fig:pd}
\end{figure}

{\em Variational Phase Diagram.}--- Having confirmed a zero hopping
Mott phase, with $n_a+n_b=1$, let us consider
itinerancy and carrier superfluidity.  Within this lowest lobe, we may
consider {\em hardcore} $a$ {\em and} $b$ bosons \footnote{Whilst this
does not affect physics {\em within} the lobe, the upper boundary is modified by the possible $b$ population.}.  A convenient approach is to augment the
variational analysis of Ref. \onlinecite{Altman:Twocomp} with a
coherent state for the light field:
\begin{equation}
\begin{aligned}
|{\mathcal V}\rangle & = |\gamma\rangle\otimes\prod_i\left[\cos\theta_i(\cos\chi_i
  a_i^\dagger
+\sin\chi_i b_i^\dagger)\right.\\
& \hspace{1.6cm}
\left. +\sin\theta_i(\cos\eta_i+\sin\eta_i b_i^\dagger a_i^\dagger)\right]
|0\rangle,
\end{aligned}
\label{varstate}
\end{equation}
where $|\gamma\rangle$ is the coherent state introduced previously, and 
$\theta,\chi,\eta,\gamma$ are to be
determined. The first term in brackets describes the Mott state, and
the second superfluidity.  For $\theta=0$ this coincides
with the variational approach for localized excitons coupled to
light~\cite{Littlewood:Models} and reproduces the previous results for
$J_\alpha=0$. More generally, (\ref{varstate}) takes into account the
effects of real hopping, involving site vacancies and
interspecies double occupation. It provides a starting point to
identify the boundaries between the Mott and superfluid regions.  We
consider spatially uniform phases, with energy density, ${\mathcal
E}\equiv\langle {\mathcal V}|H|{\mathcal V}\rangle/N$:
\begin{equation}
\begin{aligned}
{\mathcal E}& =(\tilde\epsilon_+ -\tilde\epsilon_-\cos 2\chi)
\cos^2\theta+(2\tilde\epsilon_++V)\sin^2\eta\sin^2\theta\\
& -\frac{z}{4}\left[J_a\cos^2(\chi-\eta)+J_b\sin^2(\chi+\eta)\right]
\sin^2 2\theta\\
&  +\tilde\omega\bar\gamma^2+\bar g\bar \gamma\cos^2\theta\sin2\chi,
\end{aligned}
\label{varen}
\end{equation}
where $z$ is the coordination number and
$\tilde\epsilon_\pm\equiv (\tilde\epsilon_b\pm\tilde\epsilon_a)/2$.
Minimizing on $\bar\gamma$ yields $\bar\gamma=-\bar g\cos^2\theta\sin
2\chi/2\tilde\omega$, and one may eliminate this from ${\mathcal
E}$. Exploiting symmetries one may
minimize ${\mathcal E}$ over $[0,\pi/2]$. The order parameters,
$\langle a\rangle=\frac{1}{2}\sin 2\theta \cos(\chi-\eta)$, $ \langle
b\rangle=\frac{1}{2}\sin 2\theta \sin(\chi+\eta)$, and $\langle
\psi^\dagger\psi\rangle/N=\bar\gamma^2$, yield the phase diagram in
Fig.~\ref{Fig:pd}, where $J_a=J_b=J$.
For the chosen parameters, we have four distinct phases; (i) a Mott
state with $\langle a\rangle=\langle b\rangle
=\langle\psi^\dagger\psi\rangle=0$, (ii) a superradiant Mott state
with $\langle a\rangle=\langle b\rangle=0$ and
$\langle\psi^\dagger\psi\rangle\neq 0$, (iii) a single component
superfluid with $\langle a\rangle\neq 0$ and $\langle
b\rangle=\langle\psi^\dagger\psi\rangle=0$, and (iv) a superradiant
superfluid $\langle a \rangle\neq 0$, $\langle b\rangle\neq 0$,
$\langle \psi^\dagger\psi\rangle\neq 0$. Indeed, the Hamiltonian
displays a ${\rm U}(1)\times {\rm U}(1)$ symmetry and these may be
broken independently. The phase diagram reflects this pattern of
symmetry breaking.  In particular, the superradiant Mott state
corresponds to an unbroken ${\rm U}(1)$ in the matter sector
(corresponding to a pinned density and phase fluctuations) but a
broken ${\rm U}(1)$ (or phase coherent condensate) for
photoexcitations. The expectation value of the {\em bilinear},
$\langle b^\dagger a\rangle \neq 0$, corresponds to the onset of 
coherence in the Dicke model. This novel phase may be regarded as a form of
supersolid.

In the absence of competition from other phases, the transition
between the non-superradiant insulator ($\theta=\chi=\bar\gamma=0$)
and the $a$-type superfluid ($\theta\neq 0$, $\chi=\eta=\bar\gamma=0$)
occurs when $\tilde\epsilon_a+zJ=0$. In Fig.~\ref{Fig:pd}, this is the
line $\mu_2=2(1-zJ)$.  This crosses the superradiance
boundary at a tetracritical point $(zJ^c,\mu_2^c)=(r/2,2-r)$, where
$r\equiv (1+\sqrt{5})/2$ is the Golden ratio. This follows from a
Landau expansion of (\ref{varen}); eliminating $\bar\gamma$, all the
quadratic ``mass'' terms vanish. More
generally, the phase diagram evolves with the parameters, and
the $a$-type superfluid may be replaced by the 
proximate phases \cite{BHSS:inprep}.

\begin{figure}
\begin{center}
\psfrag{label1}{(d) $b$-Atom Density}
\psfrag{label2}{(c) $a$-Atom Density}
\psfrag{label3}{(b) Reduced Photon Density}
\psfrag{label4}{(a) Total Atom Density}
\includegraphics[width=6cm,clip=true]{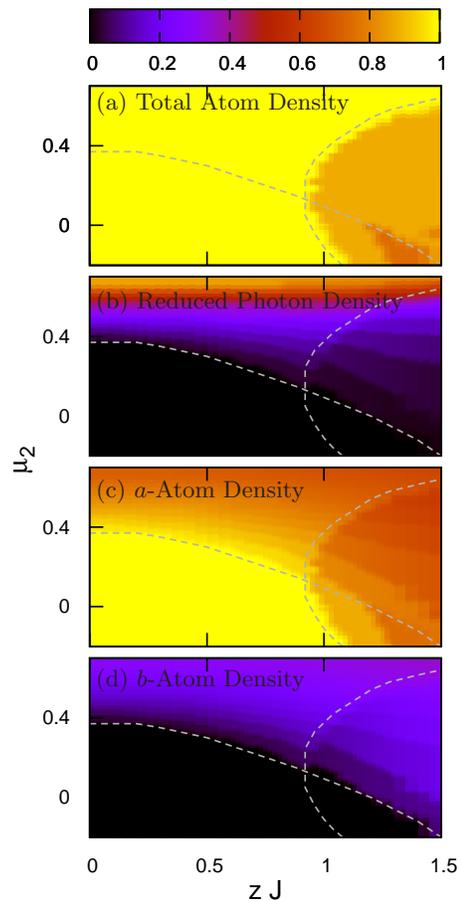}
\caption{Exact diagonalization for a 1D system with $8$ sites,
$M_\psi=16$ photons, and the parameters of
Fig.~\ref{Fig:pd}. We show (a) the total atom density and the
Mott--superfluid transition, (b) the density of photons (reduced by
a factor of two) and the onset of supperradiance, (c) the density of
$a$-atoms, and (d) the density of $b$-atoms.  The dashed lines are
a guide to the eye, and indicate the Mott--superfluid
and superradiance transitions, as determined by hand from (a)
and (b). Their intersection yields the location of the
tetracritical point.}
\label{Fig:ED}
\end{center}
\end{figure}

{\em Numerical Simulations.}--- To corroborate our findings,
we perform exact diagonalization on a 1D system of hardcore $a$ and
$b$ bosons, with $N=8$ sites and periodic boundary conditions.  The
Hilbert space is truncated to a maximum number of photons,
$M_\psi=2N=16$.  Fig.~\ref{Fig:ED} shows 
the total atom, photon, $a$-atom and $b$-atom density. The
dashed lines indicate the approximate locations of the Mott--superfluid (vertical line) and superradiance (horizontal
line) transitions, as determined from panels (a) and (b). Although an
accurate phase diagram for the thermodynamic limit is beyond the scope
of this work, the features are in excellent agreement with
Fig.~\ref{Fig:pd}.
This parallels the success of mean field theory in other
low-dimensional bosonic systems, and is remarkable given the enhanced
role of fluctuations. This may be assisted by the long range nature of
the cavity photons. The superradiance transition encompases the
superfluid and Mott phases, and yields a tetracritical point; see
(a) and (b). In addition, the region of $a$-density
over--extends that of $b$-density resulting in a pure $a$-type
superfluid; see (c) and (d).  Our simulations suggest that this
phase is stable with increasing system size~\cite{BHSS:inprep}.

{\em Discussion.}--- A feature not addressed by the present mean field
theory, but captured in Fig.~\ref{Fig:ED}, is the dispersion of the
superradiance transition with $J$; in the Mott phase, $\theta=0$, and
$J$ drops out of the variational energy (\ref{varen}).  One way to
understand this is to recast the matter contribution as $|{\mathcal
V}_{\rm M}\rangle = \prod_i(\cos\chi_i +\sin\chi_i b_i^\dagger
a_i)|\Omega\rangle$, where $|\Omega\rangle\equiv \prod_i a_i^\dagger |0\rangle$. This only
accomodates {\em local} particle-hole pairs. By analogy with the 
BCS approach to exciton
insulators~\cite{Keldysh:Kopaev}, the Mott state may be refined and
the $J$ dependence restored by incorporating momentum space
pairing~\cite{BHSS:inprep}. This connection to the BEC--BCS crossover for bosons
\cite{Koetsier:Bosecross} is reinforced by the Feshbach
resonance problem studied in the absence of a
lattice~\cite{Rad:Atmol,Romans:QPT,Zhou:PS}. Performing a
particle--hole transformation, the matter--light coupling reads
$\psi^\dagger a_i b_i$.  Aside from the global nature of the photon,
this converts $a$ and $b$ into a ``molecule'' $\psi$. At the outset
there are eight phases corresponding to condensation of $\langle
a\rangle$, $\langle b\rangle$, $\langle\psi\rangle$. Of these, only
five may survive; condensation of two variables provides an effective
field (as dictated by the coupling) which induces condensation of the
other. The band asymmetry, $\epsilon_a<\epsilon_b$, reduces this to
four, or less, depending on the parameters.  In contrast to the single
species mean field theory, this case supports an atomic superfluid, since
condensation of one carrier no longer induces a field. Moreover,
condensation may leave a ${\rm U}(1)$ symmetry intact, which allows
the coexistence of Mott insulating and phase coherent behavior.

In deriving (\ref{varen}) and the phase diagram, we are primarily
concerned with the matter-light coupling. As such, we incorporate $V$
as in Ref.~\onlinecite{Altman:Twocomp}. This gives rise to the
non-trivial phases in Fig.~\ref{Fig:pd}. However, as noted by S\"oyler
{\em et al}~\cite{Soyler:Twocomp}, analogous phases may be stabilized
in the two-component Bose--Hubbard model, {\em without} matter-light
coupling, through a more sophisticated treatment of $V$
itself. Indeed, onsite repulsive interactions, $Vn_an_b$, favor a
particle of one species and a hole of the other on the same
site. Treating this pairing in a BCS approach, one
may replace $n^a_i n^b_i$ by
$|\Delta_i|^2+(\Delta_i b_i^\dagger a_i +{\rm h.c.})$, where 
$\Delta_i\equiv \langle a_i^\dagger
b_i\rangle$, is to be determined self-consistently. This
field acts as a {\em local} ``photon'', and a similar mean field
phenomenology may ensue. Such pairing also occurs in fermionic
 models \cite{Kantian:ALE}.  Although our discussion
has focused on a single {\em global} photon, the symmetry analysis 
is more general. This is supported by
studies of the two-band Bose--Hubbard model for equal fillings and
commensurate densities \cite{Kuklov:Comm}.  We shall provide details of 
the similarities and differences of this local problem in
Ref.~\onlinecite{BHSS:inprep}. The classical limit may also be
realized in optical superlattices, where $g_ia_ib_i^\dagger$ is
tunnelling between wells, $a$ and $b$.

{\em Conclusions.}--- We have considered the impact of
photoexcitations on the Bose--Hubbard model. The phase diagram
supports a novel phase where photoexcitations condense on the
background of a Mott insulator. We have performed numerical
simulations, and highlighted connections to other problems of 
current interest.  Directions for research include the impact of fluctuations
and the nature of collective excitations.  It would also be
interesting to incorporate a finite photon wavevector. This may
stabilize inhomogeneous phases and probe incommensurate magnetism.
Recent
studies of Bose--{\em Fermi} mixtures also display a similar
phenomenology, in which superfluidity is replaced by fermionic
metalicity \cite{Sinha:Phases}.

{\em Acknowledgements.}--- We thank G. Conduit, N. Cooper and
M. K\"ohl.  We are grateful to J. Keeling for illuminating discussions.  MJB, AOS, and
BDS acknowledge EPSRC grant no. EP/E018130/1.  MH was supported by the
FWF Schr\"odinger Fellowship No.~J2583.


\end{document}